\documentclass{iau}

\usepackage{amsmath}
\usepackage{graphicx}
\usepackage{multirow}

\begin{document}

\lefttitle{Kerrison \textit{et al.}}
\righttitle{Why so young? Connecting the broadband flux and neutral gas content of AGN}

\jnlPage{1}{7}
\jnlDoiYr{2024}
\doival{xxx/xxxxx}
\volno{392}
%\jname{Title of your IAU Symposium}
\aopheadtitle{Proceedings IAU Symposium}
\editors{D.J. Pisano, Moses Mogotsi, Julia Healy, Sarah Blyth, eds.}

\title{Why so young? A curious connection between the broadband flux and neutral gas content of AGN}

\author{Emily F. Kerrison$^{1,2,3}$, Elaine M. Sadler$^{1,2,3}$,  Vanessa A. Moss$^{3,1,2}$,  Elizabeth K. Mahony$^{3,2}$ \& FLASH team}
\affiliation{$^1$Sydney Institute for Astronomy, School of Physics A28, University of Sydney, NSW 2006, Australia
\email{emily.kerrison@sydney.edu.au}}
\affiliation{$^2$ARC Centre of Excellence for All-Sky Astrophysics in 3 Dimensions (ASTRO 3D)}
\affiliation{$^3$ATNF, CSIRO, Space and Astronomy, PO Box 76, Epping, NSW 1710, Australia}

\begin{abstract}
We present here a study of the broadband spectral properties of 33 sources detected in HI absorption as part of the ASKAP-FLASH Pilot Surveys. We outline our approach to spectral classification and discuss the correlation seen between spectral shape and the detection of HI absorption. We further consider the implications of the observed correlation on the spatial distribution of the neutral gas, and on the jet-gas interactions. Our results are evaluated in the context of the forthcoming, full ASKAP-FLASH survey and other large, untargeted searches of the radio sky.
\end{abstract}


\maketitle

Although neutral hydrogen is both a key ingredient in cosmic star formation, and an excellent tracer of galaxy-scale interactions, very little is known about its distribution at intermediate redshifts. At $z \leq 1.7$ the 1215.7\,\AA\ Ly$\alpha$ line falls in the UV part of the spectrum and cannot be detected from the ground, and at $z \geq 0.4$ the 21-cm line is too faint to detect in emission with current instruments. One obvious way to fill this niche is through 21-cm absorption line studies, which allow us to detect column densities down to as low as N$_{\rm HI} \sim 10^{19}$\,cm$^{-2}$ towards bright, background radio sources \citep{Gupta2016}. 

The First Large Absorption Survey in HI (FLASH; \citealt{Allison2021}) is seeking to fill this niche, and is currently observing 24,000\,deg$^2$ of the southern sky with the Australian SKA Pathfinder (ASKAP) to search for neutral hydrogen in absorption at intermediate redshifts ($0.4 < z < 1$). This is just after cosmic noon, where we notice a marked decline in star formation rate \citep{Madau2014}, making this period critical to deepening our understanding of our Universe. This search is untargeted, and will provide us with thousands of new probes into the gaseous environments of galaxies and AGN in this under-explored redshift range. We present here a curious finding based on the broadband, radio properties of HI-detected sources in the ASKAP-FLASH Pilot Surveys. 

\section{A new, untargeted sample of HI absorbers}
Our sample is composed of 33 radio sources detected in \textrm{HI} absorption as part of ASKAP-FLASH Pilot Surveys 1 and 2. These surveys, described in \cite{Yoon2024_flash}, covered 3,000\,deg$^2$ of sky in a series of 2-hour pointings between 2019-2022, detecting over $10^4$ sources in the continuum with $S_{855.5\text{\,MHz}} > 40$\,mJy (the \textrm{HI} detection threshold of both pilot surveys). The sources in our sample were picked out from amongst these continuum sources using the Bayesian linefinder presented in \cite{Allison2012}, which identified highly significant absorption lines in each 288\,MHz-wide spectrum. These lines all have a Bayes factor greater than 30, corresponding approximately to a signal-to-noise ratio greater than 10 when integrating across the full linewidth. Our 33 radio galaxies therefore represent the strongest \textrm{HI} absorption lines in an untargeted search for neutral hydrogen at intermediate redshifts, unbiased by optical selection effects such as dust reddening and obscuration, and we expect further examination of the FLASH pilot data to reveal additional lines at lower significance levels.

From amongst these 33 detections, three were previously known from ASKAP spectral line commissioning data, and detailed studies of these objects have already been published. These are PKS\,1610-77 (an intervening line; \citealt{Sadler2020}), PKS\,1740-517 (an associated line; \citealt{Allison2015}) and PKS\,1830-210 (another intervening line first hypothesised to be from an intervening gravitational lens by \cite{Subrahmanyan1990}, and then detected in the radio by \cite{Chengalur1999}, with ASKAP data first presented in \citealt{Allison2017}). The line profiles of our sample span widths 20-200\,km/s, and in the absence of optical spectroscopy, preliminary results from the machine learning methodology of \cite{Curran2016} suggest the sample contains a mixture of associated and intervening lines, as is to be expected from an untargeted search.

\section{Broadband radio properties}

By combining many independent flux density measurements of a radio AGN on a single plot, we can create its broadband spectral energy distribution (SED). In the IR-optical-UV regime, these SEDs are commonly used as a diagnostic tool for inferring the physical parameters of a host galaxy such as stellar mass and star formation rate \citep{Pacifici2023TheTechniques}. However here, we focus on the radio frequency portion of the SED, where the AGN component typically dominates and is unimpeded by dust obscuration. Through careful modelling of this radio SED component, we can determine if a source is young, typically within ~10$^3$ years of triggering, or variable (i.e. a blazar). The youngest sources show a characteristic turnover in flux density over several decades in frequency, from which they get the name `peaked spectrum' (PS) sources \citep{ODea2021}, while blazars will exhibit stochastic changes in flux density when considering multiple epochs of observations. 

For each source in our sample, we have compiled flux densities from 80\,MHz - 100\,GHz using all large-area radio surveys currently available through the VizieR catalogue access service \citep{Ochsenbein2000}. The resulting SED of each source was then run through \textsc{RadioSED}, a Bayesian fitting framework tailored to identifying young, peaked spectrum sources, as described in \cite{Kerrison2024}. 

There is a clear over-representation of PS sources from amongst this sample, with 21/33 sources statistically favouring a peaked model. Literature estimates suggest these comprise only 10-20\% of all radio sources  \citep{ODea1998, Callingham2017}, so a fraction this high is surprising in a sample of absorbers from an untargeted search. In fact, assuming the upper limit from the literature, that 20\% of all radio galaxies are PS sources \citep{ODea1998}, the binomial probability of 21 or more HI-absorbed sources being peaked spectrum is only $\text{Pr}(\textit{N}_{\text{PS}} \ge 21) = 3 \times 10^{-8}$, and this number falls if we assume the lower population fraction from \cite{Callingham2017}. It is therefore clear that even though these \textrm{HI} absorbers are drawn from an untargeted survey, the galaxies against which we tend to detect neutral hydrogen are a specific subset that do not resemble the parent population in terms of their broadband radio characteristics.

\section{Connecting the dots}
\begin{figure}[t]
  %\centerline{\vbox to 6pc{\hbox to 10pc{}}}
  \raisebox{0.17\height}{\includegraphics[width=0.47\textwidth]{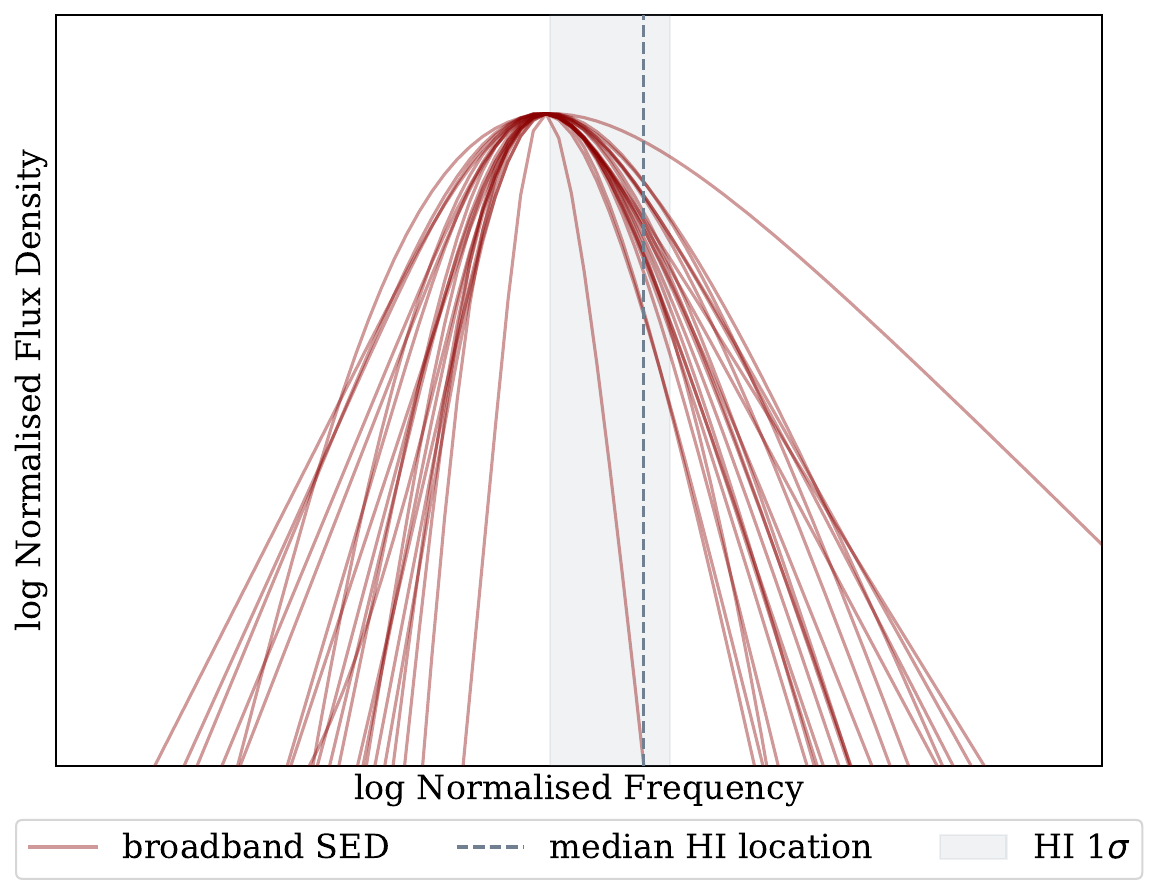}}
  \includegraphics[width=0.53\textwidth]{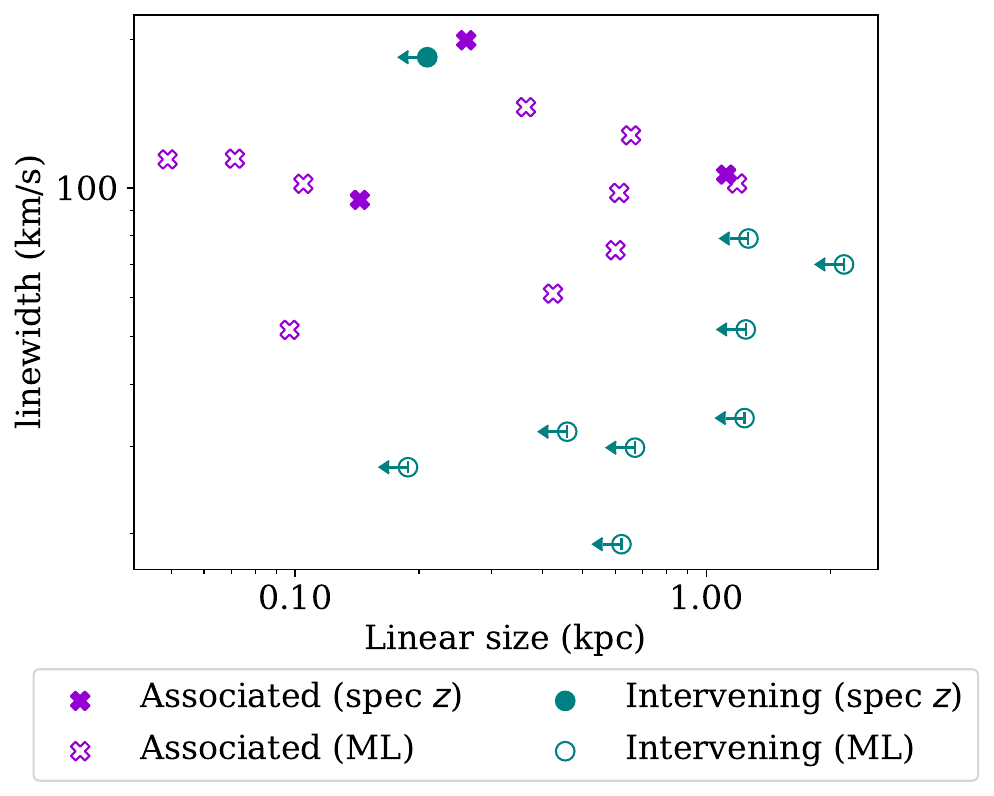}
  \caption{Left: the normalised SEDs of all 21 sources in our sample classified as peaked spectrum using \textsc{RadioSED}. The vertical dashed line indicates the median position of all \textrm{HI} lines relative to the broadband SEDs, with the shaded band indicating the spread in this position using the interquartile range. Right: linewidth of the \textrm{HI} line profiles plotted against estimates of the linear size for each source, derived from its restframe peak frequency. Points are coloured by the classification of the \textrm{HI} detection as either associated with the radio source (purple) or intervening along the line of sight (teal). The marker style further differentiates the sources into those classified by a secure, spectroscopic redshift (filled), and those classified using machine learning (empty).}
  \label{fig:normseds}
\end{figure}

Can we delve any further into the properties of this \textrm{HI}-absorbed radio population from the modest sample we have here? As a first step, we consider the apparent skew towards PS sources. To do this, we normalise all 21 PS SEDs by the location of their broadband peak and examine where the HI line falls relative to this, as shown on the left of Figure \ref{fig:normseds}. Not only is there a preference for peaked spectrum sources amongst these \textrm{HI} detections, but there is also a tendency for the rest frame peak in these sources to fall just below 1.4\,GHz, the frequency of the \textrm{HI} line, assuming all cases of absorption are associated with the radio source. In this way the presence of the \textrm{HI} line allows us to place a lower limit on the rest frame peak frequency of the PS sources, as if the absorption line is not associated but from intervening gas, the SED of the radio galaxy will move further to the right in the rest frame of the source.

Going one step further, we can also use the relative positions of the \textrm{HI} line and the broadband spectral turnover to place limits on the linear sizes of the sources in our sample. It is well known that the peak of a PS source correlates closely with its linear size \citep[e.g.][]{ODea1997ConstraintsSources, Jeyakumar2016}, so we here combine our estimates of rest frame peak frequencies with the empirical relation from \cite{Jeyakumar2016} (especially Figure 3 of that work) to determine an approximate linear size for the PS sources in our sample. These estimates are plotted against the linewidth of the \textrm{HI} absorption profiles in the right of Figure \ref{fig:normseds}, where we have further divided the PS sources of our sample into those identified as associated (purple) and intervening (teal), where the latter will only provide upper limits on the linear size of the background radio source, as well as into those securely identified as associated/intervening using optical spectroscopy (filled markers), and those identified in a preliminary application of machine learning (empty markers).

The width of the absorption line profile is not strongly correlated with the linear size of the radio source in this sample. However, we note the preliminary nature of the machine learning classifications, and we stress that larger samples will be required to more fully explore any connections between source structure and profile shape. A simpler observation with regards to this sample is that these sources all have estimated linear sizes between 0.05 – 2.2 kpc, with a mean size of $0.6\pm0.5$\,kpc. This aligns with the high \textrm{HI} detection fraction seen at these sizes in inhomogeneous samples and supports the idea of a `resonance’ between the size of the absorbing screen and the radio source, as discussed in \cite{Curran2013}.

\section{Future work}

Our investigation into the broadband radio spectral properties of the 33 \textrm{HI} detections from the untargeted ASKAP-FLASH Pilot Surveys is summarised here. We have observed a statistically significant preference for \textrm{HI} absorption to be detected towards young, peaked spectrum radio sources. Not only that, but the location of the spectral peaks in these sources suggest that they all have linear sizes less than $\sim2$\,kpc, matching the preference for compact sources seen in inhomogeneous, targeted samples of \textrm{HI} detections from the literature. It will be particularly interesting to see how these trends develop as we move into full survey mode, and the untargeted ASKAP-FLASH sample grows to hundreds, then thousands of absorbers. Alongside this growing sample, high-resolution VLBI imaging, combined with optical spectroscopy of the host galaxies will be key to better understanding the nature of the neutral gas distribution at these intermediate redshifts.

\bibliographystyle{iaulike_updated}
\bibliography{references}

\end{document}